\newcommand{\pcmq}{\mbox{cm$^{-2}$}}
\newcommand{\psec}{\mbox{s$^{-1}$}}
\newcommand{\funit}{\mbox{ph \pcmq \psec}}
\newcommand{\eunit}{\mbox{erg \pcmq \psec}}

\def\deg{\ensuremath{^\circ}}

\documentclass[11pt,twoside]{article}
\usepackage{asp2006}
\usepackage{epsf}
\usepackage{graphicx}
\usepackage{lscape}
\def\edcomment#1{\iffalse\marginpar{\raggedright\sl#1\/}\else\relax\fi}
\reversemarginpar

\begin{document}
\title{Fermi Observations of the Large Magellanic Cloud}
\author{J. Kn\"odlseder for the Fermi LAT collaboration}
\affil{Centre d'\'Etude Spatiale des Rayonnements, CNRS/Universit\'e de Toulouse,
PO Box 44346, 31028 Toulouse Cedex 4}

\begin{abstract}
We report on observations of the Large Magellanic Cloud with the {\em Fermi Gamma-Ray Space 
Telescope}. 
The LMC is clearly detected with the Large Area Telescope (LAT) and for the first time the emission 
is spatially well resolved in gamma-rays. 
Our observations reveal that the bulk of the gamma-ray emission arises from the 30 Doradus 
region. 
We discuss this result in light of the massive star populations that are hosted in this area and 
address implications for cosmic-ray physics. 
We conclude by exploring the scientific potential of the ongoing Fermi observations on the study 
of high-energy phenomena in massive stars.
\end{abstract}

\vspace{-0.5cm}
\section{Introduction}

Since the early days of high-energy gamma-ray astronomy it has become clear that the
gamma-ray flux received at Earth is dominated by emission from the Galactic disk
\citep{clark68}.
This emission can be well understood in terms of cosmic-ray interactions with the
interstellar medium \citep{strong07}.
At energies $\ga100$~MeV, the generation of diffuse gamma-ray emission is dominated
by the decay of $\pi^0$ produced in collisions between cosmic-ray nuclei and interstellar
medium nuclei.
Ultimately, the study of this hadronic gamma-ray emission may provide hints on the still 
mysterious origin of the galactic cosmic-rays.
However, the interpretation of the galactic diffuse gamma-ray emission is complicated by
the fact that a large number and variety of individual sources contribute  along the line of sight
to the observed emission, thus blurring the link between individual cosmic-ray acceleration sites 
and observed gamma-ray signatures in our Galaxy.

Gamma rays from cosmic-ray interactions are also expected from nearby galaxies, and
indeed, the EGRET telescope aboard the Compton Gamma-Ray Observatory ({\em CGRO})
has for the first time detected gamma-ray emission from the Large Magellanic Cloud 
(LMC) \citep{sreekumar92}.
The LMC is an excellent target for studying the link between cosmic-ray acceleration
and gamma-ray emission since this galaxy is nearby (bringing the sources fluxes in
reach of modern gamma-ray telescopes) and since the system is nearly seen face-on
(avoiding the superposition of sources along the line of sight that hampers studies in
our own Galaxy).
In addition, the LMC is rather active, housing many supernova remnants, bubbles
and superbubbles and massive star forming regions that are all potential sites
of cosmic-ray acceleration \citep{biermann04,cesarsky83,binns07}.

The Large Area Telescope (LAT) aboard  the {\em Fermi Gamma-Ray Space Telescope} 
(FGST) provides now the capabilities to study diffuse gamma-ray emission from nearby
galaxies in depth, and of the LMC in particular \citep{digel00,weidenspointner07}.
We report here on the initial analysis of observations taken in the course of the
first year's all-sky survey by the LAT .

\section{Observations}

The LAT is the primary instrument on the FGST satellite which has been launched 
from Cape Canaveral on June 11th, 2008.
The LAT is an imaging, wide field-of-view, high-energy gamma-ray telescope, covering
the energy range from below 20 MeV to more than 300 GeV.
The LAT is a pair-conversion telescope with a precision tracker made of a stack of 18
x,y silicon tracking planes and a calorimeter made of 96 CsI(Tl) crystals.
The tracker array is covered by a segmented anticoincidence shield allowing for the
rejection of charged particle backgrounds.

The LAT has a large $\sim2.5$~sr field of view, and compared to earlier gamma-ray 
missions, has 
a large effective area ($>7000$~cm$^2$ on axis at $\sim1$~GeV for the event
selection used in this paper),
improved angular resolution ($\sim0.5\deg$ 68\% containment radius at 1 GeV) and
low dead time ($\sim25$~$\mu$s per event).
The $1\sigma$ energy resolution in the 100 MeV - 10 GeV energy range is better 
than $\sim10$\%.
A detailed description of the instrument is given by \citet{atwood09}.
The on-orbit instrument calibration is presented by \citet{abdo09a}.

The data used in this work covers the period August 8th 2008 -- April 24th 2009
and amounts to 211.7 days of continuous sky survey observations.
During this period a total exposure of 
$\sim2.3 \times 10^{10}$~cm~s$^2$
(at 1 GeV) has been obtained for the LMC region.

\subsection{Data preparation}

The data analysis presented in this paper has been performed using the
\break{\tt ScienceTools} version {\tt v9r11} and the instrument response functions
{\tt P6\_V3}.
We collected all data obtained within the period August 8th 2008 -- April 24th 2009
and applied the {\em diffuse event class} filter that has been designed to minimize 
contamination by instrumental background while retaining a substantial fraction
of the signal.
As has been pointed out by \citet{atwood09}, any harsher event cut would not
significantly improve the signal-to-noise ratio.

We further excluded from the data all periods where the spacecraft has entered
the South Atlantic Anomaly (SAA) and for which the spacecraft z-axis points
more than $47\deg$ away from the zenith direction (the zenith direction being
defined by the vector running from the Earth center through the spacecraft).
While the SAA cut excludes periods of particular large instrumental background
from the analysis, the latter cut excludes periods where the Earth enters the field of 
view.
Furthermore, to minimize contamination from Earth albedo photons we exclude 
photons with zenith angles above $105\deg$ from the analysis.
We further restricted the analysis to photon energies above 200~MeV where our
current knowledge of the instrument response implies systematic uncertainties
that are smaller than $\sim10\%$ and where the redistribution of photons in energy
due to incomplete energy measurements becomes negligible.

\subsection{Morphology}

\begin{figure}[!t]
\centering
\includegraphics[width=10cm]{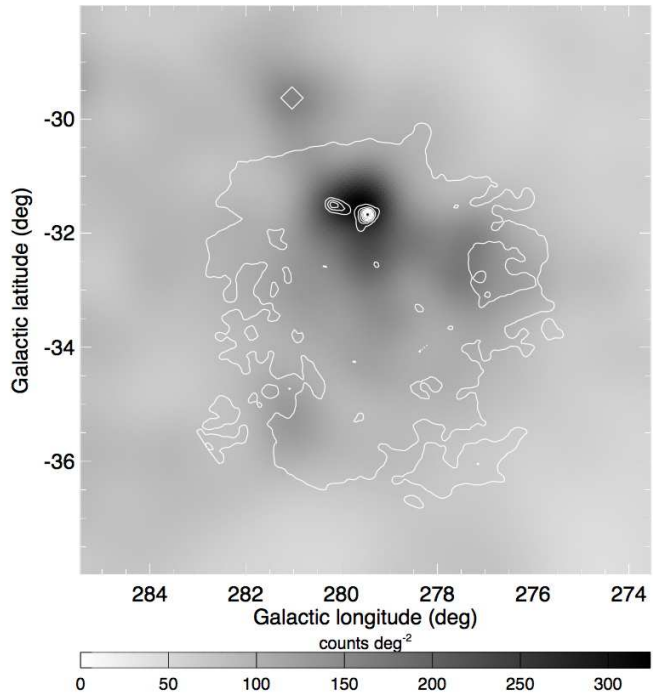}
\caption{
Preliminary adaptively smoothed (s.n.r.~$=10$) {\em Fermi}/LAT counts map of a 
$10\deg\times10\deg$ 
large region centered on the LMC for the energy range 200 MeV - 100 GeV (greyscale).
The contours show the extinction map of \citet{schlegel98} as tracer of the total gas 
column density in the LMC.
Ten linearly spaced contour levels are plotted.
The diamond in the north-east of the image designates the location of the blazar
CRATES~J060106-703606 \citep{healey2007}
that contributes at a low level to the gamma-ray emission in this area.
\label{fig:image}
}
\end{figure}

To illustrate the distribution of observed gamma-ray photons in the LMC region
we show in Fig.~\ref{fig:image} a counts-map of the area.
The arrival directions of observed photons in the 200 MeV - 100 GeV energy range
have been binned into $3' \times 3'$ large pixels covering a $10\deg\times10\deg$ 
large area around the position $(l,b) = (279.5\deg, -33.0\deg)$.
The binned map has then been smoothed using a 2D adaptive Gaussian kernel smoothing
technique \citep{ebeling06} to remove Poissonian noise that arises from the relatively
small number of counts that have been registered.
The signal-to-noise ratio (s.n.r.) has been set to 10 to reduce statistical noise variations
to below $\la10\%$ in the image.

We overlay as contours on the {\em Fermi}/LAT counts map the extinction map of 
\citet{schlegel98} as tracer of the total gas column density in the LMC.
To first order the extinction scales linearly with total gas column density, and we
chose 10 linearly spaced contours to allow the reader to visually appreciate the
distribution of gas column densities in the LMC.
Obviously, a substantial fraction of the gas is found in a small area in the north
of the LMC, at roughly $(l,b) \sim (279.5\deg, -31.5\deg)$, which coincides with
the 30~Doradus star forming region.

\begin{figure}[!t]
\centering
\includegraphics[width=10cm]{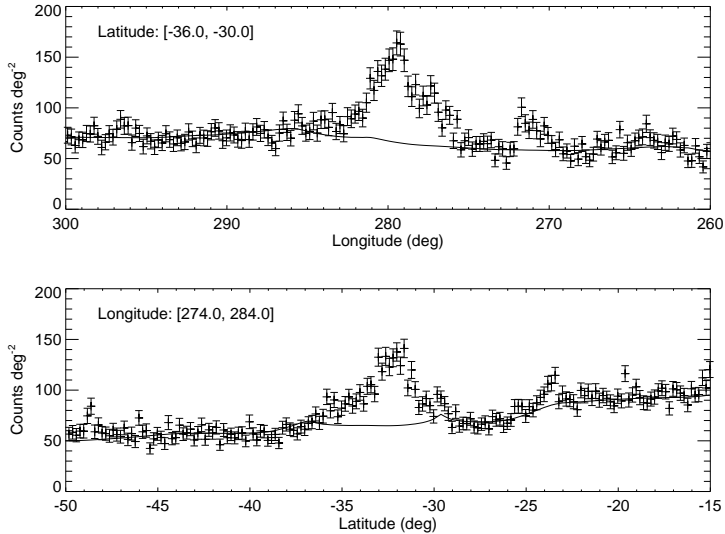}
\caption{
Preliminary longitude (top) and latitude (bottom) photon intensity profiles of the LMC region for the
energy range 200 MeV - 100 GeV.
The solid line indicates the expected contributions from diffuse galactic emission, diffuse
extragalactic emission, instrumental background and the blazar CRATES~J060106-703606 at
$(l,b) = (281.04\deg,-29.63\deg)$ in this area of the sky.
\label{fig:profile}
}
\end{figure}

The high-energy gamma-ray photons that are observed from the LMC also peak in 
this area. 
The photon intensity in the 30~Dor region exceeds $\sim300$ counts/deg$^2$
while in most of the remaining regions of the LMC is remains below $\sim120$
counts/deg$^2$ (the background rate around the LMC is around
$\sim50-70$ counts/deg$^2$).
The excess near 30~Dor is also clearly seen in the longitude and latitude profiles
of the photon intensity observed by LAT that is shown in Fig.~\ref{fig:profile}.

Within a rectangular box covering galactic longitudes $274\deg \le l \le 284\deg$ and
galactic latitudes $-36\deg \le b \le -30\deg$ we find a total number of
$\sim1800$ counts within the energy range 200 MeV - 100 GeV
above the expected contributions from galactic diffuse emission,
extragalactic diffuse emission, instrumental background, and the 
blazar CRATES~J060106-703606.
These background contributions have been estimated by fitting spatial and spectral 
templates of their emission components together with a spatial template for the 
LMC emission to the data.
Galactic diffuse emission has been modelled spatially and spectroscopically
using the GALPROP model \citep{strong07} version {\tt 54\_59Xvarh7S}\footnote{
Available from the website: http://galprop.stanford.edu}
while for the combination of extragalactic diffuse emission and residual instrumental
background we assume an isotropic emission with a power law spectral distribution.
CRATES~J060106-703606 is modeled as a point source at
$(l,b) = (281.04\deg,-29.63\deg)$
with a power-law spectral distribution.
For the LMC we use the extinction map of \citet{schlegel98} as spatial template
from which we subtract a pedestal level of $0.07^{\rm m}$ from all pixels and
for which we set all pixels outside a radius of $4\deg$ around
$(l,b) = (279.65\deg,-33.34\deg)$
to zero in order to extract the LMC emission.
As spectral model we assume a power law for the LMC.

To describe the morphology of the high-energy gamma-ray emission from the LMC
we first fit a point source with free position and flux on top of the background
model\footnote{From now on we call the combination of the GALPROP model, the 
isotropic model and the CRATES~J060106-703606 point source the 
{\em background model} of our analysis. The free parameters of this background
model are the normalization of the GALPROP model, the intensity and spectral slope
of the isotropic component, and the flux and spectral slope of the 
CRATES~J060106-703606 point source.} to our data.
This results in a best-fitting point-source position of
$(l,b) = (279.58\deg, -31.72\deg)$
with a statistical 95\% confidence error radius of
$0.09\deg$ (the systematic position uncertainty is estimated to less
than $0.02\deg$).
We note that this position is close to that of R~136,
the central star cluster of 30~Dor, which is located at 
$(l,b) = (279.47\deg, -31.67\deg)$,
i.e. at an angular distance of $0.11\deg$ from our best-fitted point-source
location.

The detection significance of the LMC can be estimated using the so-called
{\em Test Statistics} (TS) which is defined as twice the difference
between the log-likelihood $L_1$ that is obtained by fitting the LMC model
on top of the background model to the data, and
the log-likelihood $L_0$ that is obtained by fitting the background model only,
i.e. ${\rm TS} = 2(L_1 - L_0)$.
Under the hypothesis that our model satisfactorily explains the {\em Fermi}/LAT data,
TS follows a $\chi^2_p$ distribution with $p$ degrees of freedom,
where $p$ is the number of free parameters in the LMC model 
\citep{cash79}.
In the particular case of a point source with free position, flux and spectral index
we have $p=4$ and the measured TS of 869.1 corresponds to a
significance of $29.8 \sigma$.

As next step we replace the point source model by an extended source model 
which we implement as axisymmetric 2D Gaussian shape with variable angular
size $\sigma$.
In addition to the size we again fitted the position, flux and power law spectral 
index of the source.
This results in a best-fitting source position of
$(l,b) = (279.5\deg, -32.2\deg)$ (with a 95\% confidence radius of $0.1\deg$),
and source extent of
$\sigma = 1.0 \pm 0.1\deg$.
The TS amounts to 1088.5 which is larger by 219.4 than the value obtained 
for the point-source model.
Since we added one additional parameter (the source extent $\sigma$) with respect
to the point-source model we obtain the significance of the source extension from the 
$\chi^2_1$ distribution to $14.8\sigma$.

Alternatively to the geometrical models we also compare the {\em Fermi}/LAT
data to various spatial templates that trace the interstellar matter distributions
in the LMC.
For neutral hydrogen (H~I) we use the aperture synthesis and multibeam data that
\citet{kim2005} have combined from ATCA and Parkes observations.
For molecular hydrogen we use CO observations of the LMC obtained with
the NANTEN telescope \citep{yamaguchi01}.
We further used the extinction map of \citet{schlegel98} (SFD) as tracer of the total gas
column density and compare also our data to the 100 $\mu$m IRIS map that has
been obtained by reprocessing the IRAS survey data \citep{mivilledeschenes05}.
The results of this comparison are summarized together with that of the geometrical
models in Table \ref{tab:models}.

The best fits are obtained for the SFD extinction map and the IRAS 100 $\mu$m infrared
map which give TS values of $1179.6$ and $1179.1$, respectively.
For 2 free parameters (the total flux in the map and the spectral index) this corresponds to
a detection significance of $34.5\sigma$.
An almost equally good fit is obtained using the neutral hydrogen map.
Fitting instead the CO map to the LAT data provides a rather poor fit, suggesting that
the gamma-ray morphology differs from that of molecular gas in the LMC.
Fitting the H~I and CO maps together to the data confirms this result since the fit
attributes $97\%$ of the total flux to the H~I  component.
Correspondingly, the TS increase with respect to fitting the H~I gas map alone is 
also negligible.

The H~I/SFD/IRIS~100~$\mu$m maps fit the data considerably better than a single
point source, adding further evidence that the observed high-energy gamma-ray
emission is extended in nature.
Furthermore, the 2D Gaussian source model cannot reach the fit that is obtained by those
tracer maps, suggesting that the emission morphology is more complex than a
single Gaussian shape.

\begin{table}[!t]
\caption{Comparison of maximum likelihood model fitting results (see text for a
description of the models). Column 1 gives the model used to the fit the LMC
data, column 2 gives the TS value of the fit, and column 3 specifies the number of
free parameters of the LMC model.\label{tab:models}}
\smallskip
\begin{center}
{\small
\begin{tabular}{lrc}
\tableline
\noalign{\smallskip}
LMC model & TS & Parameters \\
\noalign{\smallskip}
\tableline
\noalign{\smallskip}
Point source & 869.1 & 4 \\
2D Gaussian source & 1088.5 & 5\\
H~I gas map & 1173.4 & 2 \\
CO gas map & 932.2 & 2 \\
H~I + CO gas maps & 1176.1 & 4 \\
SFD extinction map & 1179.6 & 2 \\
IRIS 100 $\mu$m infrared map & 1179.1 & 2 \\
\noalign{\smallskip}
\tableline\
\end{tabular}
}
\end{center}
\end{table}

\subsection{Spectrum}
 
\begin{figure}[!t]
\centering
\includegraphics[width=10cm]{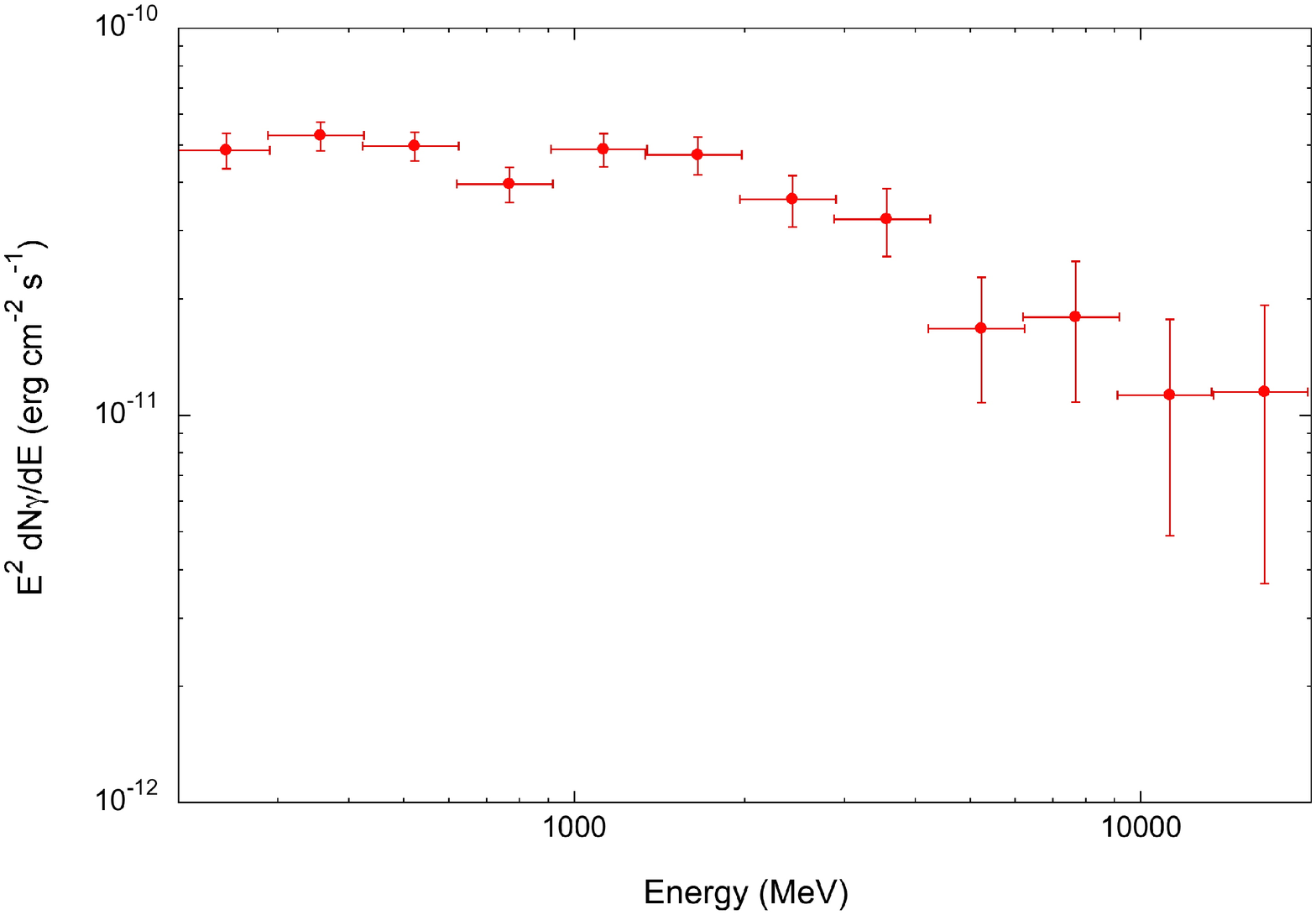}
\caption{
Preliminary spectrum of the LMC obtained by fitting the extinction map of \citet{schlegel98} in 12
logarithmically-spaced energy bins covering the energy range 200 MeV - 20 GeV to the
{\em Fermi}/LAT data.
Errors are statistical only.
\label{fig:spectrum}
}
\end{figure}

Using the extinction map of \citet{schlegel98} (i.e. our best fitting spatial template of the 
high-energy emission) we extract a spectrum of the LMC by fitting the data in
12 logarithmically-spaced energy bins covering the energy range 200 MeV - 20 GeV.
Above 20 GeV, photons from the LMC become too sparse in our actual data set to allow
for meaningful spectral points to be derived.

Figure~\ref{fig:spectrum} shows the LMC spectrum that has been obtained by
this method.
Our analysis indicates a spectral steepening of the emission with increasing energy,
suggesting that a simple power law is an inadequate description of the data.
We confirm this trend by fitting the data using a broken power law instead of a simple 
power law.
This results in an improvement of the TS by $10.1$ corresponding to a significance
of $2.7\sigma$ ($p=2$) of the spectral steepening.
Fitting alternatively an exponentially cutoff power law improves the TS by $7.8$ with
respect to the simple power law, corresponding to a significance of $2.8\sigma$ ($p=1$)
of the spectral cutoff.
Integrating the broken power law or the exponentially cutoff power law model over the
energy range 100~MeV -- 500~GeV gives identical
photon fluxes of $(3.1 \pm 0.2) \times 10^{-7}$ \funit\ and an
energy fluxes of $(2.0 \pm 0.1) \times 10^{-10}$ \eunit\
for the LMC.
The systematic uncertainty in these flux measurements amounts to $\sim10\%$.

\section{Discussion and conclusions}

\citet{sreekumar92} reported the first detection of the LMC in $>100$~MeV gamma rays
based on 4 weeks of data collected with the EGRET telescope aboard {\em CGRO}.
Due to EGRET's limited angular resolution and the weak emission detected from
the LMC, details of the spatial structure of the galaxy could not have been resolved.
However, it has been obvious from EGRET data that the LMC was an extended gamma-ray 
source.

{\em Fermi}/LAT allows now for the first time to clearly resolve the gamma-ray
emission of the LMC and to attribute the emission maximum to the 30~Dor
star forming region.
While this coincidence could be taken as a hint for an enhanced cosmic-ray
density in 30~Dor with respect to the rest of the galaxy, we note that a substantial
fraction of the interstellar gas of the LMC is also confined to the 30~Dor area.
Consequently, the target density for cosmic-ray interactions is greatly enhanced in
this region which implies a corresponding enhancement of the gamma-ray luminosity.
Whether the data do also support an enhanced cosmic-ray density in 30~Dor
with respect to the rest of the galaxy needs a more detailed analysis of the observations.

The rather poor fit of the CO map to the LAT data suggests that the overall distribution 
of gamma-ray emission differs from that of molecular hydrogen.
The distribution of neutral hydrogen fits the data considerably better and the 
combined fit of H~I and CO maps indicates that any contribution to the gamma-ray
emission that is correlated to the molecular gas is at best marginal.
This agrees well with expectations since the gas budget of the LMC is largely
(90-95\%) dominated by neutral hydrogen \citep{fukui99}.
Consequently we are presently unable to determine the CO-to-H$_2$ conversion 
factor, X$_{\rm CO}$, from our LMC data.

\citet{fichtel91} performed a detailed modelling of the cosmic-ray distribution in the 
LMC and predicted an integrated $>100$~MeV photon flux of
$(2.3 \pm 0.4) \times 10^{-7}$ \funit\ for the galaxy.
\citet{pavlidou01} predicted an integrated $>100$~MeV photon flux of
$1.1\times 10^{-7}$ \funit\ based on estimates of the LMC supernova rate and 
total gas densities.
Our observed flux of
$(3.1 \pm 0.2) \times 10^{-7}$ \funit\ 
falls at the hide side of these estimates, yet given the uncertainties in the
models the agreement can be judged satisfactorily.

Further studies of the LMC with {\em Fermi}/LAT will now concentrate on the 
spectral analysis of the data, with particular emphasize on variations of the spectral
shape throughout the galaxy.
Thanks to the excellent sensitivity and angular resolution of the LAT, this is the
first time that such studies become possible.
And other nearby galaxies await their detection, such as the Small Magellanic
Cloud or the Andromeda Galaxy (M31).
Both should be in reach of {\em Fermi} and the comparative study of their diffuse
gamma-ray emission should help to understand the impact of the environment
and metallicity on the physics of cosmic rays.


\end{document}